\documentclass[%
 reprint,
nofootinbib,
 amsmath,amssymb,
 aps,
]{revtex4-2}

\usepackage{graphicx}
\usepackage{dcolumn}
\usepackage{bm}

\usepackage{subfigure,graphicx,epsfig,amsmath,amsfonts,amssymb,xcolor,slashed,ulem,multirow}

\usepackage{relsize}
\usepackage{slashed}
\usepackage{color}
\usepackage{ulem}
\usepackage{tabu}

\usepackage{amsmath}
\usepackage{amssymb}
\usepackage{amsthm}
\usepackage{mathrsfs}
\usepackage{graphicx}
\usepackage{epstopdf}
\usepackage{fancyhdr}
\usepackage{array}
\usepackage[all]{xy}
\usepackage{eufrak}
\usepackage{euscript}
\usepackage{enumerate}
\usepackage{slashed}
\usepackage{hyperref}
\usepackage{subfigure} 
\usepackage{epstopdf} 

\usepackage{hyperref}
\hypersetup{pdftex,colorlinks=true,linkcolor=blue,citecolor=blue,menucolor=black,urlcolor=blue,filecolor=blue}

\newcommand{\be}{\begin{equation}}
\newcommand{\ee}{\end{equation}}
\newcommand{\ba}{\begin{eqnarray}}
\newcommand{\ea}{\end{eqnarray}}
\newcommand{\nn}{\nonumber}

\begin{document}
\title{Production of pentaquarks with hidden charm and double strangeness in $\Xi_b$ and $\Omega_b$ decays}

\date{\today}

\author{E.~Oset}
\email{oset@ific.uv.es}
\affiliation{Departamento de F\'{\i}sica Te\'orica and IFIC, Centro Mixto Universidad de
Valencia-CSIC Institutos de Investigaci\'on de Paterna, Aptdo.22085,
46071 Valencia, Spain}

\author{L. Roca}
\email[]{luisroca@um.es}
\affiliation{Departamento de F\'isica, Universidad de Murcia, E-30100 Murcia, Spain}

\author{M.~Whitehead}
\email{Mark.Whitehead@glasgow.ac.uk}
\affiliation{School of Physics and Astronomy, University of Glasgow, Glasgow, United Kingdom}

\begin{abstract}

Recently, several pentaquark states $P_{css}$, with global flavor  $\bar c c s s n$, have been predicted within a theoretical framework based on unitary coupled channels.
We  study theoretically the feasibility to observe the $P_{css}$ with 
$I(J^P)=\frac{1}{2}(\frac{1}{2}^-)$ in the decays $\Xi_b^0\to \eta \eta_c \Xi^0$ and $\Omega_b^-\to K^- \eta_c \Xi^0$. Indeed, within the model, the $\eta_c \Xi^0$ channel is the lowest mass pseudoscalar-baryon channel to which this pentaquark state couples, thus we can expect to observe its signal in the $\eta_c \Xi^0$  invariant mass distribution of the mentioned decays.
We identify the dominant weak decay processes
and then implement the hadronization into the different meson-baryon channels in
the final state, linked by flavor symmetry. The dominant meson-baryon final state interaction is then implemented to  generate the full amplitude, implicitly accounting for the dynamical emergence of the pentaquark states.
We obtain a clear Breit-Wigner-like resonant signal in the spectrum of the  $\Omega_b^-$ decay, exceeding that in the $\Xi_b^0$ decay by two to three orders of magnitude. In the case of the latter decay, the resonant state would manifest as a significant dip in the spectrum.
We study the feasibility of searching for these $b$-hadron decay modes and analysing their resonant components using the current and future data samples from the LHCb experiment.

\end{abstract}

\maketitle


\section{Introduction}

Pentaquark research has become highly active in the past decade, spurred by notable experimental findings from the LHCb collaboration, which reported hidden charm pentaquark candidates such as $P_c(4380)$, $P_c(4312)$, $P_c(4440)$, $P_c(4457)$, $P_c(4337)$ without strangeness \cite{LHCb:2015yax,LHCb:2019kea,LHCb:2021chn}, as well as containing a strange quark,
$P_{cs}(4459)$ \cite{LHCb:2020jpq} and $P_{cs}(4338)$ \cite{LHCb:2022ogu}. Several theoretical models, in particular those based on meson-baryon molecular interpretations, have sought to elucidate their nature
(see Refs.~\cite{Chen:2016qju,Lebed:2016hpi,Esposito:2016noz,Guo:2017jvc,Ali:2017jda,Liu:2019zoy,Chen:2022asf} for reviews on the subject). Several states were indeed predicted  \cite{Wu:2010jy,Wu:2010vk,Wang:2011rga,Yang:2011wz,Wu:2012md,Xiao:2013yca,Li:2014gra,Chen:2015loa,Karliner:2015ina} before their observation by LHCb.

A logical progression would be to explore the potential existence of hidden-charm pentaquarks with strangeness $S=-2$, denoted as $P_{css}$, for which there is currently no experimental evidence. 
This contrasts with theoretical predictions of models relying on implementing unitarity in coupled channels, based on potentials derived from t-channel vector meson exchange \cite{Marse-Valera:2022khy,Roca:2024nsi}, which have predicted several $P_{css}$ states in the range about 4500-4700~MeV.
Other theoretical  approaches are based on the meson exchange model \cite{Wang:2020bjt}, sum rules \cite{Azizi:2021pbh} and quark models \cite{Anisovich:2015zqa,Ortega:2022uyu}.

The states in Refs.~\cite{Marse-Valera:2022khy,Roca:2024nsi,Wang:2020bjt} emerge as generated from the intricate nonlinear dynamics involved in the unitarization of the meson-baryon scattering amplitudes. In particular, 
the lowest mass pole predicted in Ref.~\cite{Roca:2024nsi}, was found in the 
pseudoscalar-meson--baryon interaction with quantum numbers $I(J^P)=\frac{1}{2}(\frac{1}{2}^-)$ and flavor  $\bar c c s s n$,
which was associated to a molecular pentaquark state with mass about 4500~MeV and width of the order of 10~MeV, predominantly coupling to $\bar D \Omega_c$ and $\bar D_s\Xi'_c$, and less strongly to $\eta_c \Xi $.

This theoretical backdrop prompts exploration of potential experimental reactions aimed at observing such a state.
This endevour can draw inspiration from the success story of the pentaquark with  hidden charm and single strangeness, $P_{cs}$. Indeed this state was anticipated by coupled channel unitary models \cite{Wu:2010jy,Wu:2010vk,Xiao:2019gjd},
and the $\Xi_b^-\to J/\Psi K^-\Lambda$ decay was proposed in Ref.~\cite{Chen:2015sxa} to observe its signal. This reaction was later employed by the LHCb collaboration to experimentally observe this pentaquark for the first time \cite{LHCb:2020jpq}.

Encouraged by this successful example, in the present work we study the possibility to see the $P_{css}$ pentaquark with $I(J^P)=\frac{1}{2}(\frac{1}{2}^-)$ in the $\eta_c \Xi^0$ spectrum in the $\Xi_b^0\to \eta \eta_c \Xi^0$ and $\Omega_b^-\to K^- \eta_c \Xi^0$ decays. The choice of $\eta_c \Xi $ 
for the observation of the $P_{css}$ state is justified since the $\bar D \Omega_c$ and 
 $\bar D_s\Xi'_c$ channels are closed for decay.

We also discuss if these reactions could be attainable in an experimental facility such as LHCb.

\section{Formalism}

\subsection{Summary of the unitary coupled channel model}

We first provide a brief summary of the model from Ref.~\cite{Roca:2024nsi} for generating the pentaquark state with flavor $\bar c c s s n$ and  $I(J^P)=\frac{1}{2}(\frac{1}{2}^-)$, further details can be found in that reference. The model follows the techniques
of the chiral unitary approach
\cite{Kaiser:1995eg,Oset:1997it,Oller:1998zr,Oller:2000fj,Dobado:1996ps,Oller:1998hw,Nieves:2001wt}
 to sum up the coupled
channels $\eta_c \Xi $, $\bar D_s\Xi'_c$ and $\bar D \Omega_c$, implementing unitarity, and starting from kernel potentials
based on t-channel vector meson exchange mechanisms.
The required vertices are obtained using a formalism  based on heavy
quark spin symmetry and the local hidden
gauge approach \cite{Bando:1984ej,Bando:1987br,Birse:1996hd,Meissner:1987ge,Nagahiro:2008cv}  properly extrapolated to the charm sector \cite{Molina:2009ct,Molina:2010tx}.

Within this model, the S-wave pseudoscalar-baryon ($PB$) tree level potentials are given by \cite{Roca:2024nsi}
\begin{eqnarray}
  \label{eq:def_Vij}
     V_{ij}=g^2 C_{ij}(p^0+p'^{0}) \, ,
\end{eqnarray}
where the subindices denote specific pseudoscalar-baryon channels, $p^0(p'^0)$ are the on-shell center of mass energy of the initial(final) meson, $g$ is a coupling constant defined in \cite{Roca:2024nsi}, and the coefficients $C_{ij}=C_{ji}$ are given in Table~\ref{Tab:CijPBVPu}.
\begin{table}[h!]
  \begin{center}
    \begin{tabular}{c|cccc}
        \hline\\ [-0.30cm]
         & $\eta_c\Xi$ &  $\bar{D}_s\Xi_c^\prime$ & $\bar{D}\Omega_c$\\
        \hline \\ [-0.2cm]
        $\eta_c\Xi$ & $0$ & $\frac{1}{\sqrt{6}m^2_{D^*_s}}$ & $-\frac{1}{\sqrt{3} m^2_{D^*}}$ \\ 
        $\bar{D}_s\Xi_c^\prime$ & &$\frac{1}{m^2_\phi}-\frac{1}{m^2_{J/\psi}}$ & $\frac{\sqrt{2}}{m^2_{K^*}}$ \\
        $\bar{D}\Omega_c$ & & & $-\frac{1}{m^2_{J/\psi}}$ \\
    \end{tabular}
  \end{center}
\caption{$C_{ij}$ coefficients of the $PB$ interaction in the $\bar c c s s n$ sector.}    
\label{Tab:CijPBVPu}
\end{table}
Note the dependence on the inverse of the squared masses of the vector mesons exchanged in the t-channel.

\begin{figure}[h]
     \centering
     \includegraphics[width=.99\linewidth]{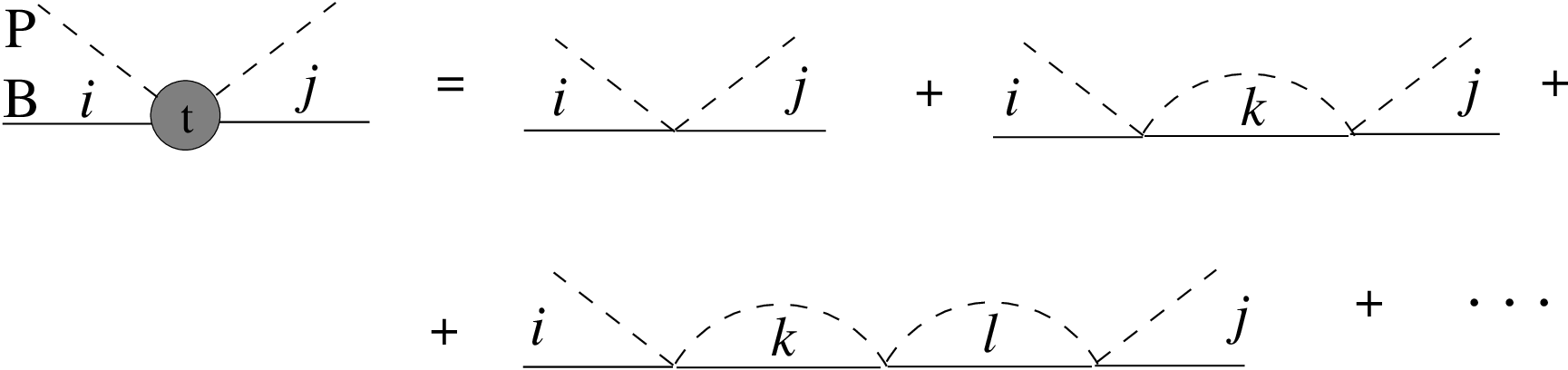}
    \caption{Unitarization of the pseudoscalar-baryon interaction.}
    \label{fig:BS}
\end{figure}

Building upon the  potentials of Eq.~\eqref{eq:def_Vij}, within the framework  of the coupled channels unitary approach, exact unitarity can be incorporated into the meson-baryon interaction. To this aim, we use the Bethe-Salpeter equation
\begin{equation} t=(1-VG)^{-1}V \ ,
\label{eq:BS}
\end{equation} where $G$ represents a diagonal matrix containing the meson-baryon loop functions.
It is worth interpreting Eq.~(\ref{eq:BS}) as the coupled channel resummation shown in Fig.~\ref{fig:BS}, where each intermediate step includes all of the possible $PB$ states listed in Table~\ref{Tab:CijPBVPu}.

The loop function, $G_l$, for the meson-baryon channel $l$, is regularized by means of a three-momentum cutoff, $\Lambda$:
\begin{eqnarray}
        G_l(\sqrt{s})  =  \int_{q< \Lambda} &&\frac{d^3\vec q}{(2\pi)^3}\frac{1}{2\omega_l({q})}\frac{M_l}{E_l({q})} \nonumber\\
                &\cdot&  \frac{1}{\sqrt{s}-\omega_l({ q}) - E_l(q)+i\epsilon}\,,
          \label{eq_Gl}
\end{eqnarray}	  
with  $q=|\vec q|$, $\omega_l(q)=\sqrt{m^2_l+q^2}$, $E_l=\sqrt{M^2_l+q^2}$, and $m_l(M_l)$ the mass of the meson(baryon).
The value of the cutoff stands as the primary source of uncertainty within the model, naturally ranging between   $\Lambda=600$~MeV and $800$~MeV  \cite{Roca:2024nsi}.
A potential issue with the cutoff method in the heavy hadron sector arises from the presence of a pole in the integrand of Eq.~\eqref{eq_Gl} for $\sqrt{s}$ in the range $[m+M,\sqrt{\Lambda^2+m^2}+\sqrt{\Lambda^2+M^2}]$. This means that near $\sqrt{s}\sim\sqrt{\Lambda^2+m^2}+\sqrt{\Lambda^2+M^2}$, the result for the real part of the loop function may not be reliable, as the integrand above the pole cannot adequately compensate the sharp increase before  the pole in the proper evaluation of the principal value. 
This concern with the cutoff method for heavy hadrons is also discussed in Refs.~\cite{Wu:2010rv,Xiao:2013jla,Feijoo:2022rxf}.
In the present work, it only affects the $\eta_c \Xi^0$ channel for $\Lambda\sim600$~MeV since, for this channel,
$\sqrt{\Lambda^2+m^2}+\sqrt{\Lambda^2+M^2}=4489$~MeV, which is close to the mass of the predicted pentaquark state, around $\sim4500$~MeV. 
However, since at energies close to the pentaquark mass we are significantly above the $\eta_c \Xi^0$ threshold,  the real part of its loop function must be small, with the imaginary part dominating the behavior. Therefore, for $\Lambda\sim 600$~MeV,  we consider only the imaginary part in the $\eta_c \Xi^0$ loop function as a suitable approximation. This procedure is commonly used when one has a situation like this \cite{Dai:2020yfu}.

The high nonlinear dynamics involved in Eq.~\eqref{eq:BS} leads to the emergence of poles in the $t_{ij}$ scattering amplitudes on the second Riemann sheet. From the position of these poles, the value of the mass and width of the generated resonances can be inferred. In the present case, for $I(J^P)=\frac{1}{2}(\frac{1}{2}^-)$, a pole was found in 
Ref.~\cite{Roca:2024nsi} corresponding to a state of mass 4535~MeV and a width of 9~MeV for $\Lambda=600$~MeV, and $M=4479$~MeV, $\Gamma=12$~MeV, for $\Lambda=800$~MeV. The difference between the results for both cutoff values is to be considered as a measure of the uncertainty in our model. 
Additionally, from the residue of the amplitudes at the poles, the couplings of the generated resonances to the different channels can be obtained. We found in Ref.~\cite{Roca:2024nsi}
 that this pentaquark couples mostly to $\bar D \Omega_c$ and 
 $\bar D_s\Xi'_c$ and very weakly to  $\eta_c \Xi $ (explicit values of the couplings can be found in Table IV of Ref.~\cite{Roca:2024nsi}). However, since $\eta_c \Xi $ is the only open channel for the generated state, it is the only source of the width of the pentaquark. 

The generated pentaquark described in this section is to be understood as dynamically arising from the coupled-channel effects, and the present study aims to investigate its production in $\Xi_b^0\to \eta \eta_c \Xi^0$ and $\Omega_b^-\to K^- \eta_c \Xi^0$ decays.

\subsection{Formalism for $\Xi_b^0\to \eta \eta_c \Xi^0$}

In this section we explain
the mechanism for the $\Xi_b^0\to \eta \eta_c \Xi^0$ process for which we follow an analogous formalism to that used in previous studies such as the calculation of
$\Xi_b^-\to J/\Psi K^- \Lambda$ \cite{Chen:2015sxa},
$\Lambda_c^+\to\pi^+ MB$ \cite{Miyahara:2015cja} (with $MB$ a meson-baryon pair), or $\Lambda_b\to J/\Psi \Lambda(1405)$ \cite{Roca:2015tea},
which are based on a methodology initially proposed in \cite{MartinezTorres:2009uk}.

\begin{figure}[h]
\centering
\includegraphics[width=.7\linewidth]{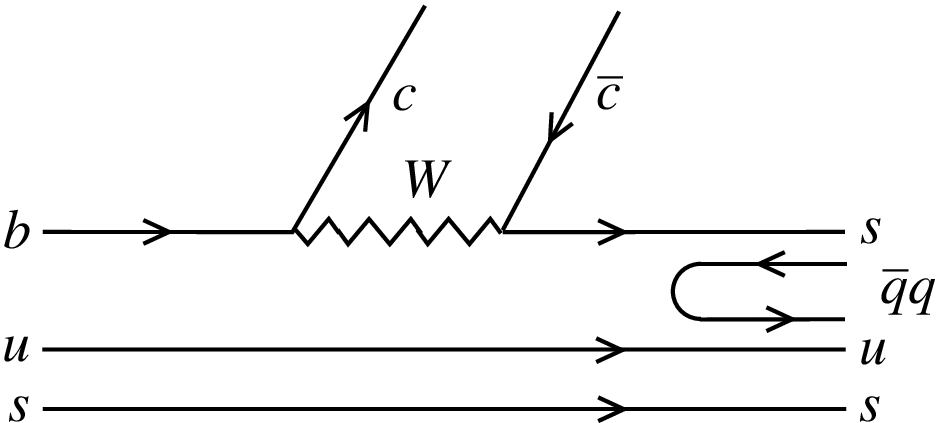}
\caption{Dominant quark diagram for $\Xi_b^0\to \eta_c MB$ decay, prior to the final state interaction.}
\label{fig:diagram_tree_Xib}
\end{figure}

The reaction mechanism, prior to the final state interaction (FSI), is diagrammatically depicted in Fig.~\ref{fig:diagram_tree_Xib}, and can be understood by dividing it into three parts. 

\noindent\textbf{ Weak decay}: First the $b$ quark within the $\Xi_b^0$ state undergoes a weak transition to form a $c\bar c $ pair and an $s$-quark, as illustrated in the left part of Fig.~\ref{fig:diagram_tree_Xib}, resulting in a final three-quark system $sus$. 
This transition is characterized by
the matrix elements of the Cabibbo-Kobayashi-Maskawa
(CKM) matrix $V^{\phantom{*}}_{cb}V^*_{cs}$. This process is commonly referred to as an internal W-emission diagram. 
External W-emission, which is color favored, does not work for this decay but it is
possible in $\Omega_b^-\to K^- \eta_c \Xi^0$, as will be discussed later.
Further discussion and justification of the suppression of alternative topologies for  weak mechanisms at the quark level can be found in Refs.~\cite{Miyahara:2015cja,Miyahara:2016yyh}.

\noindent\textbf{Hadronization:}
The $c\bar c $ pair combines to create the $\eta_c$ meson, while the virtual $sus$ three-quark state undergoes hadronization to form a meson-baryon pair. This hadronization proceeds via the generation of a $\bar q q$ pair with vacuum quantum numbers, in a philosophy in the line to the  $^3P_0$ model \cite{Micu:1968mk,LeYaouanc:1972vsx}.
In the present process, the formation of a $\bar q q$ pair between the lower $u$ and $s$ lines is not possible. Indeed, initially strong correlation between the $u$ and $s$ quarks of the original $\Xi_b$ state is assumed, with the quarks moving independently within a potential well. Since the $\Xi_b$ ($J^P=1/2^+$) resides in the ground state of the three-quark $(usb)$ configuration, the relative angular momenta between the individual quarks are all zero. Following the weak transition, but before the hadronization process, the three-quark state $(sus)$ must be in a p-wave to produce a pentaquark with $J^P=1/2^-$. As the initial $u$ and $s$ quarks are regarded as spectators and initially possess $L=0$, the only possible scenario is for the upper $s$ quark to carry an angular momentum of $L=1$. Since the final-state mesons and baryons are in their respective ground states and in S-wave in relation to each other, all angular momenta in the final state are zero. Therefore, the $\bar q q$ pair is constrained to be produced solely between the upper $s$ and $u$ quarks, as depicted in Fig.~\ref{fig:diagram_tree_Xib}.

To evaluate the scattering amplitude for the process depicted in Fig.~\ref{fig:diagram_tree_Xib}, let us start by considering the wave function of the initial $\Xi_b^0$ in flavor space  \cite{Capstick:1986ter,Roberts:2007ni,Wang:2022aga} :

\begin{align*}
|\Xi_b^0\rangle=\frac{1}{\sqrt{2}}|b(us-su)\rangle\,,
\end{align*}
which is antisymmetric in the lighter quarks.
After the weak process, but before the hadronization, the three-quark state turns into
\begin{align*}
\frac{1}{\sqrt{2}}|s(us-su)\rangle\,,
\end{align*}
which preserves the same antisymmetric flavor and spin correlation given that the initial $u$ and $s$ quarks are considered as spectators.

To incorporate the hadronization, we introduce the $\bar q q $ pairs, leading to the final quark flavor state:
\begin{align*}
|H\rangle&\equiv \frac{1}{\sqrt{2}}\eta_c|s\,(\bar u u +\bar d d +\bar s s+\bar c c)\,(us-su)\rangle\\
         &=\frac{1}{\sqrt{2}}\eta_c \sum_{i=1}^4{|P_{3i}q_i(us-su)}\rangle\,,
\end{align*}
where
\begin{equation}
q\equiv \left(\begin{array}{c}u\\d\\s\\c\end{array}\right)\,\text{~~and~~~}
P\equiv q\bar q^\tau=\left(\begin{array}{cccc}
u\bar u & u\bar d & u\bar s & u\bar c\\
d\bar u & d\bar d & d\bar s & d\bar c\\
s\bar u & s\bar d & s\bar s & s\bar c\\
c\bar u & c\bar d & c\bar s & c\bar c\\
			      \end{array}\right)\,,
\label{eq:P1}
\end{equation} 
which corresponds to  the quark-antiquark representation of the
pseudoscalar meson matrix
\begin{eqnarray}
P=
\left(\begin{array}{cccc} 
              \frac{\pi^0}{\sqrt{2}}  + \frac{\eta}{\sqrt{3}}+\frac{\eta'}{\sqrt{6}}& \pi^+ & K^+ & \bar D^0\\
              \pi^-& -\frac{1}{\sqrt{2}}\pi^0 + \frac{\eta}{\sqrt{3}}+ \frac{\eta'}{\sqrt{6}}& K^0 & D^-\\
              K^-& \bar{K}^0 & -\frac{\eta}{\sqrt{3}}+ \frac{2\eta'}{\sqrt{6}} & D_s^-\\
	      D^0 & D^+ & D_s^+ & \eta_c
      \end{array}
\right)\,,
\label{eq:P2}
\end{eqnarray}
where for the $\eta$ and $\eta'$ we have assumed the usual physical mixing
between the singlet and octet $SU(3)$ states \cite{Bramon:1992kr}.

The hadronized state $|H\rangle$ can now be written as
\begin{align}
|H\rangle= &\frac{1}{\sqrt{2}} \eta_c\bigg( K^- u(us-su)+\bar K^0 d(us-su)\nn \\
&+\frac{1}{\sqrt{3}}\left(-\eta+\sqrt{2}\eta'\right)s(us-su)
+ D_s^-c(us-su)\bigg),
\label{eq:Hhadrons}
\end{align}
where we omit the {\it bra-ket} notation for simplicity.
The $\eta_c\eta s(us-su)$ term is the only one in Eq.~\eqref{eq:Hhadrons} that exhibits an overlap with one of the channels\footnote{Note that the last term in Eq.~\eqref{eq:Hhadrons} would generate the state $D_s^-\Xi_c^+$, not the $D_s^-\Xi_c^{'+}$ of Table~\ref{Tab:CijPBVPu}.}
responsible for generating the $P_{css}$ in our model, namely $\eta_c \Xi^0$.
Actually, this term overlaps with the antisymmetric component of the $\Xi^0$ baryon. Indeed, the $\Xi^0$ state comprises a combination of both symmetric and antisymmetric flavor wave functions:
\begin{eqnarray}
\Xi^0=\frac{1}{\sqrt{2}}\left(\phi_{MS}\chi_{MS}+\phi_{MA}\chi_{MA}\right),
\label{eq:X0sym}
\end{eqnarray}
where $\phi_{MS(MA)}$ stands for the mixed-symmetric (antisymmetric) flavor wave function upon permutation of quarks 2 and 3, while $\chi$ serves a similar purpose for spin \cite{close_book,Wang:2022aga}.
Specifically, the antisymmetric flavor component\footnote{\label{foot} There is a different sign convention in the $\Xi^0$ flavor state with respect to Ref.~\cite{close_book}, as explained in Refs.~\cite{Pavao:2017cpt,Miyahara:2016yyh}.} required in the present case is \cite{close_book}:
\begin{eqnarray*}
\Xi^0\sim-\frac{1}{2}s(us-su),
\end{eqnarray*}
thus the amplitude  for the process of Fig.~\ref{fig:diagram_tree_Xib} leading to $\eta_c \eta \Xi^0$ is 
\begin{align}
{\cal M}= {\cal C} \langle H|\eta_c \eta \Xi^0\rangle=\,\frac{{\cal C}}{\sqrt{6}},
\label{eq:MXib}
\end{align}
where ${\cal C}$ is a global factor incorporating  the
weak interaction process, including matrix elements of the two 
$\gamma^\mu(1-\gamma_5)$ $Wqq$ vertices, essentially given by the corresponding CKM matrix elements, as well as the $W$ propagator. Additionally, $\cal{C}$ encompasses dynamical factors of the matrix elements for
the hadronization process, except for the flavor structure which is explicitly taken into account in ${\cal M}$.

\begin{figure}[h]
\centering
\includegraphics[width=1\linewidth]{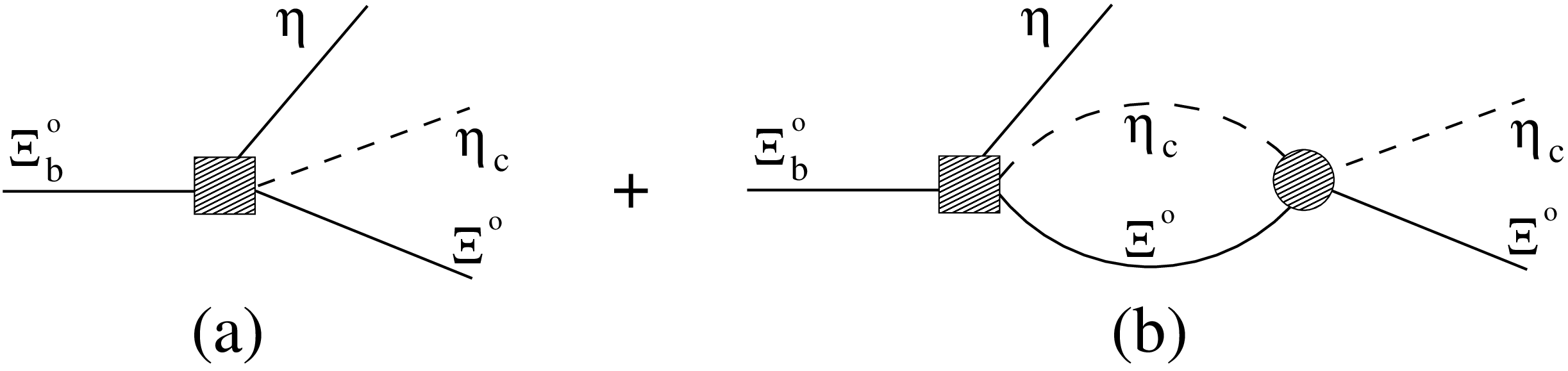}
\caption{Final-state interaction of the meson-baryon pair.
The square denotes the production mechanism of the $\eta\eta_c\Xi^0$ 
as depicted in Fig.~\ref{fig:diagram_tree_Xib}, and the circle  the coupled channel meson-baryon scattering 
matrix $t_{11}$ in Eq.~\eqref{eq:BS} and Fig.~\ref{fig:BS}.}
\label{fig:diag_FSI}
\end{figure}

\noindent\textbf{Final state interaction}:
The final, yet crucial, stage in producing the pentaquark,  $P_{css}$, involves accounting for the final state interaction of the $\eta_c\Xi^0$ pair, depicted in Fig.~\ref{fig:diag_FSI}(b). This interaction  has to be added to the tree level of Fig.~\ref{fig:diagram_tree_Xib}, more simply represented by Fig.~\ref{fig:diag_FSI}(a). 
In Fig.~\ref{fig:diag_FSI} the square represents the tree level production of Fig.~\ref{fig:diagram_tree_Xib}, while the circle represents the meson-baryon scattering 
matrix $t_{11}$ from Eq.~\eqref{eq:BS} and Fig.~\ref{fig:BS}.

The decay amplitude for the 
$\Xi_b^0\to \eta \eta_c \Xi^0$ process, represented in Fig.~\ref{fig:diag_FSI}, can thus be written as

\begin{align}\label{eqn:fullamplitude}
\mathcal{M}(M_{\rm inv})={\cal C}\frac{1}{\sqrt{6}}\left( 1+G_{\eta_c \Xi^0}(M_{\rm inv})\,t_{\eta_c \Xi^0,\eta_c \Xi^0}(M_{\rm inv}) \right)\,,
\end{align}
where $G$ denotes the same meson-baryon loop function as in Eq.~\eqref{eq_Gl}, $t_{\eta_c \Xi^0,\eta_c \Xi^0}$ is the $\eta_c \Xi^0 \to \eta_c \Xi^0$ unitarized scattering amplitude (element $t_{11}$ in Eq.~\eqref{eq:BS}), and $M_{\rm inv}\equiv M_{\eta_c \Xi^0}$
is the invariant mass of the $\eta_c \Xi^0$ system in the final state. 
While it seems that the $1/\sqrt{6}$ in Eq.~\eqref{eqn:fullamplitude} could have been absorbed in the global factor ${\cal C}$, it is important to retain it in this form in order to compare the relative strength with the $\Omega_b^-\to K^- \eta_c \Xi^0$ process,
which we will describe in the next section.

 Finally, the $\eta_c \Xi^0$ invariant mass distribution  can be expressed as:
\begin{align}\label{eqn:dGammadM}
\frac{d\Gamma}{dM_{\rm inv}}(M_{\rm inv})
=\frac{1}{(2\pi)^3}\frac{M_{\Xi^0}}{M_{\Xi_b^0}}p_{\eta} p_{\eta_c}\left|\mathcal{M}(M_{\rm inv})\right|^2\,,
\end{align}
where $p_{\eta}$ and $p_{\eta_c}$ denote the modulus of the 
three-momentum
of the $\eta$ in the $\Xi_b^0$ rest-frame, and the modulus of the center-of-mass
three-momentum of the $\eta_c$ in the final $\eta_c \Xi^0$ system, respectively.

\subsection{Formalism for $\Omega_b^-\to K^- \eta_c \Xi^0$}

In this section, we focus solely on the differences and particularities  of the $\Omega_b^-\to K^- \eta_c \Xi^0$ decay with respect to the process $\Xi_b^0\to \eta \eta_c \Xi^0$, as its formalism parallels that of the preceding one.

\begin{figure}[h]
\centering
\includegraphics[width=.99\linewidth]{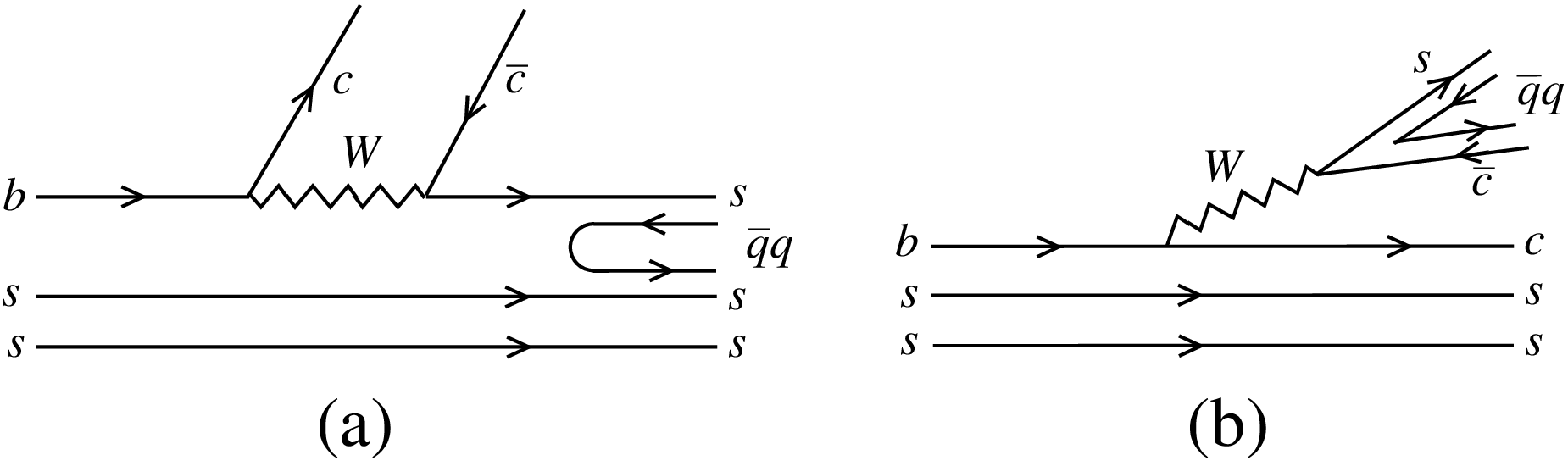}
\caption{Dominant quark diagrams for $\Omega_b^-\to K^- \eta_c \Xi^0$ decay, prior to the final state interaction.}
\label{fig:diagram_tree_Omegab}
\end{figure}

In Fig.~\ref{fig:diagram_tree_Omegab}(a), we see that the internal W-emission is analogous to Fig.~\ref{fig:diagram_tree_Xib}, but now the external W-emission mechanism is also possible, Fig.~\ref{fig:diagram_tree_Omegab}(b). In the latter case, the $\bar q q$ production of the hadronization process must occur between the $\bar c$ and $s$ quarks in order to have the opportunity to produce any of the final states of 
Table~\ref{Tab:CijPBVPu}.

Considering that now the $\Omega_b^-$ flavor state is just
\begin{align*}
|\Omega_b^-\rangle=|bss\rangle\,,
\end{align*}
which is symmetric under the exchange of the 2nd and 3rd quarks,
the resulting quark flavor state after the hadronization shown in Fig.~\ref{fig:diagram_tree_Omegab}(a) is given by
\begin{align*}
|H\rangle&\equiv \eta_c|s\,(\bar u u +\bar d d +\bar s s+\bar c c)\,ss\rangle\\
         &=\eta_c \sum_{i=1}^4{|P_{3i}q_iss}\rangle\,,
\end{align*}
which, in terms of pseudoscalar mesons, is
\begin{align}
|H\rangle&= \eta_c\bigg( K^- uss+\bar K^0 dss \nn\\
&+\frac{1}{\sqrt{3}}\left(-\eta+\sqrt{2}\eta'\bigg)sss
+ D_s^- css \right).
\end{align}
Now the only overlap with one of the possible channels of Table~\ref{Tab:CijPBVPu} comes from the first term, which goes as
 $K^-\eta_c \Xi^0$. More specifically, we now require the mixed-symmetric part of 
Eq.~\eqref{eq:X0sym}, which is \cite{close_book} (with the extra minus sign of footnote \ref{foot} incorporated)
\begin{eqnarray*}
\Xi^0\sim \frac{1}{\sqrt{2}}\frac{1}{\sqrt{6}}(sus+ssu-2uss),
\end{eqnarray*}
and thus, the amplitude  for the process of Fig.~\ref{fig:diagram_tree_Omegab}(a) (internal emission) is  
\begin{align}
{\cal M}_{ie}= {\cal C} \langle H|K^-\eta_c  \Xi^0\rangle=\,-\frac{{\cal C}}{\sqrt{3}}.
\label{eq:MieC}
\end{align}
Note that the global coefficient ${\cal C}$ is the same as in Eq.~\eqref{eq:MXib}, but Eq.~\eqref{eq:MieC} differs from Eq.~\eqref{eq:MXib} in the flavor coefficient.

The implementation of the final state interaction onto the internal emission diagram of Fig.~\ref{fig:diagram_tree_Omegab}(a) is represented by  the same mechanism as in Fig.~\ref{fig:diag_FSI}, but with the replacement of $\Xi_b^0$ by $\Omega_b^-$ and $\eta$ by $K^-$. The resulting amplitude for this process is:
\begin{align}\label{eqn:fullamplitudeMie}
\mathcal{M}_{ie}(M_{\rm inv})=-{\cal C}\frac{1}{\sqrt{3}}\left( 1+G_{\eta_c \Xi^0}(M_{\rm inv})\,t_{\eta_c \Xi^0,\eta_c \Xi^0}(M_{\rm inv}) \right)\,.
\end{align}

\begin{figure}[h]
\centering
\includegraphics[width=.6\linewidth]{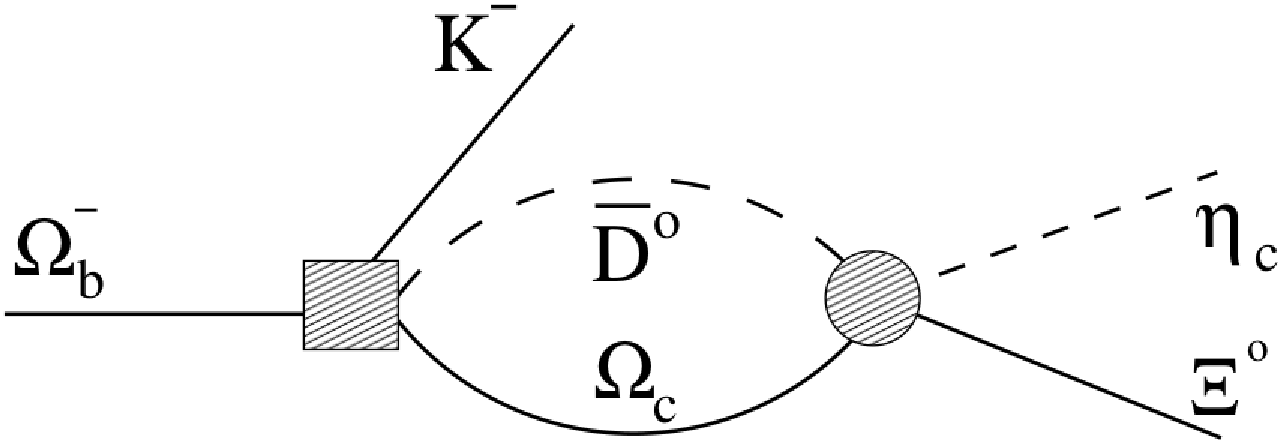}
\caption{Final-state interaction for the external emission process. 
The square denotes the production mechanism of the $K^-\eta_c\Xi^0$ 
as depicted in Fig.~\ref{fig:diagram_tree_Omegab}(b), and the circle  the coupled channel meson-baryon scattering 
matrix $t_{31}$ in Eq.~\eqref{eq:BS}.}
\label{fig:diag_FSI_ee}
\end{figure}

For the external emission process, Fig.~\ref{fig:diagram_tree_Omegab}(b), the $css$ quarks form an $\Omega_c$ and the $\bar q q$ pairs are inserted between the $s$ and $\bar c$ quarks. This leads to the final state in flavor space being
\begin{align}
|H\rangle&\equiv \Omega_c|s\,(\bar u u +\bar d d +\bar s s+\bar c c)\,\bar c\rangle\nn\\
&=\Omega_c \sum_{i=1}^4 P_{3i}P_{i4}\,=\Omega_c P^2_{34}\nn\\
&
=\Omega_c\bigg(K^-\bar D^0+\bar K^0 D^- -\frac{1}{\sqrt{3}}\eta D_s^-\nn\\
&+\sqrt{\frac{2}{3}}\eta'D_s^-+D_s^-\eta_c\bigg),
\label{eq:HOm}
\end{align}
from where we can see that although there is no tree level decay into $K^- \eta_c \Xi^0$, such a final state can be produced via coupled channels from the transition 
$\bar D^0 \Omega_c\to  \eta_c \Xi^0$, as illustrated in Fig.~\ref{fig:diag_FSI_ee}.

Therefore, the amplitude  for the external emission process of the $\Omega_b^-\to K^- \eta_c \Xi^0$  decay is 
\begin{align}\label{eqn:fullamplitudeMee}
\mathcal{M}_{ee}(M_{\rm inv})=N_c{\cal C}G_{\bar D^0\Omega_c}(M_{\rm inv})\,t_{\bar D^0\Omega_c,\eta_c \Xi^0}(M_{\rm inv}) \,.
\end{align}
Note that the external W-emission is enhanced by a factor of the number of colors, $N_c=3$, compared to the internal W-emission. This arises from the need to sum over the three possible quark colors in the hadronization
process in Fig.~\ref{fig:diagram_tree_Omegab}(b), unlike in  Fig.~\ref{fig:diagram_tree_Omegab}(a) where the color is fixed in the final meson. There is indeed an ambiguity in the relative sign when compared to the  internal emission process, Eq.~\eqref{eqn:fullamplitudeMie}, which affects the interference between these two terms. However, as we will see in the results section, the $\Omega_b^-\to K^- \eta_c \Xi^0$  decay spectrum is largely determined by the external-emission mechanism, rendering this ambiguity in sign rather inconsequential.

Finally, the expression for the invariant mass distribution is the same as in Eq.~\eqref{eqn:dGammadM} but substituting the $\Xi_b^0$ by $\Omega_b^-$ and the $\eta$ by $K^-$, and the amplitude is now $\mathcal{M}=\mathcal{M}_{ie}+\mathcal{M}_{ee}$.


\section{Results}

\begin{figure}[h]
     \centering
     \subfigure[]{\includegraphics[width=.95\linewidth]{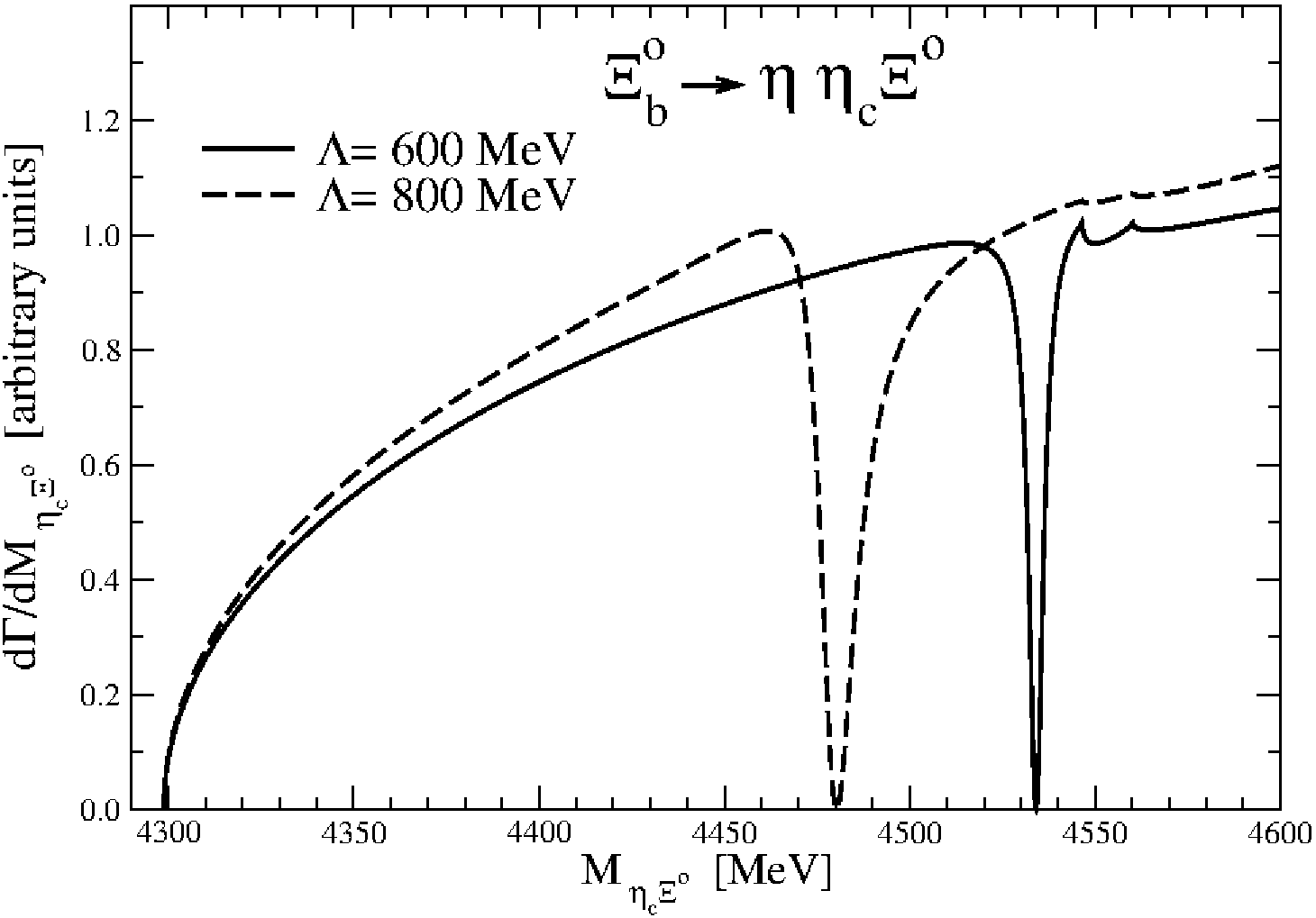}} \\
     \subfigure[]{\includegraphics[width=.95\linewidth]{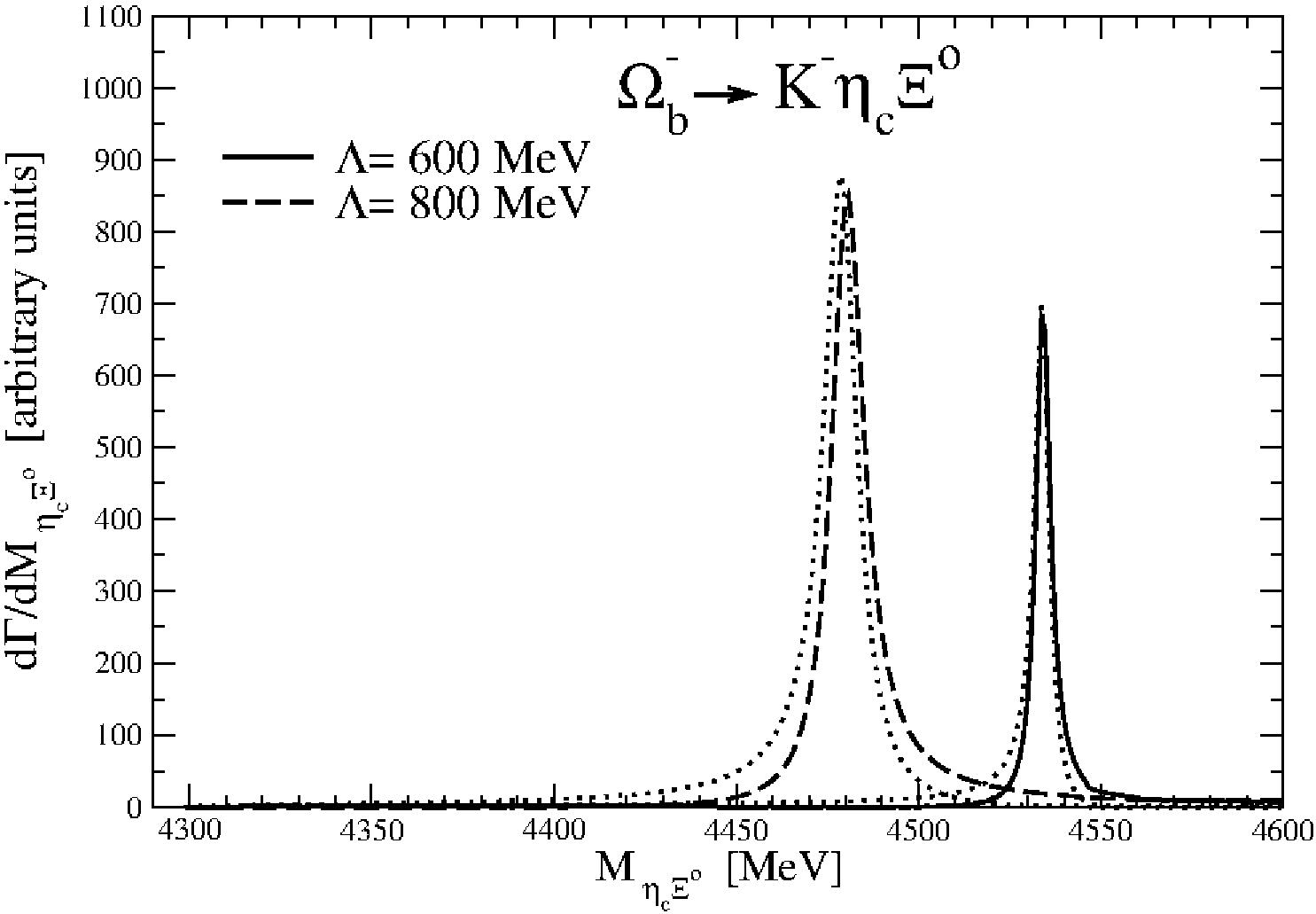}} \\
    \caption{The $\eta_c \Xi^0$  invariant mass distributions for $\Xi_b^0\to \eta \eta_c \Xi^0$ and $\Omega_b^-\to K^- \eta_c \Xi^0$ decays, for two different values of the meson-baryon regularization cutoff, $\Lambda$. The dotted lines in the lower panel represent the analogous results changing the sign of Eq.~\eqref{eqn:fullamplitudeMee}.
}
\label{fig:results1}
\end{figure}

In Fig.~\ref{fig:results1}, we show the $\eta_c \Xi^0$  invariant mass distributions for the $\Xi_b^0\to \eta \eta_c \Xi^0$ and $\Omega_b^-\to K^- \eta_c \Xi^0$ decays. In each plot, the solid line represents the result obtained with a cutoff of $\Lambda=600$~MeV, while the dashed line corresponds to $\Lambda=800$~MeV. The dotted lines in the plot for the $\Omega_b^-\to K^- \eta_c \Xi^0$ decay represent the results obtained by changing the sign in Eq.~\eqref{eqn:fullamplitudeMee} to account for the ambiguity in its relative sign compared to Eq.~\eqref{eqn:fullamplitudeMie}, as explained in the previous section. The plots clearly show that the uncertainty resulting from this effect is very small, primarily as a result of the considerable predominance of external emission relative to internal one in this decay.

Although there is a global arbitrary normalization factor, the relative strengths between the different reactions reflect genuine theoretical predictions.
An immediately striking observation from the figures is the significant difference in intensity and shape between the distributions of both decays, calling for a deeper examination of the underlying reasons.
Regarding the large difference in strength in favor of the $\Omega_b^-$ decay, it is a consequence of the interplay of various contributing factors. Firstly, the  $\Omega_b^-$ decay is predominantly governed by the external W-emission mechanism. This amplitude enjoys a $N_c=3$ enhancement, see Eq.~\eqref{eqn:fullamplitudeMee}, with respect to the 
internal W-emission amplitude in $\Xi_b^0$ decay, Eq.~\eqref{eqn:fullamplitude}. In addition, the latter one is suppressed by a $1/\sqrt{6}$ factor. When these amplitudes are squared to compute the spectrum, the $\Omega_b^-$ decay ends up with a magnitude approximately 50 times larger, solely due to numerical flavor and color coefficients.
\begin{figure}[h]
     \centering
     \includegraphics[width=.99\linewidth]{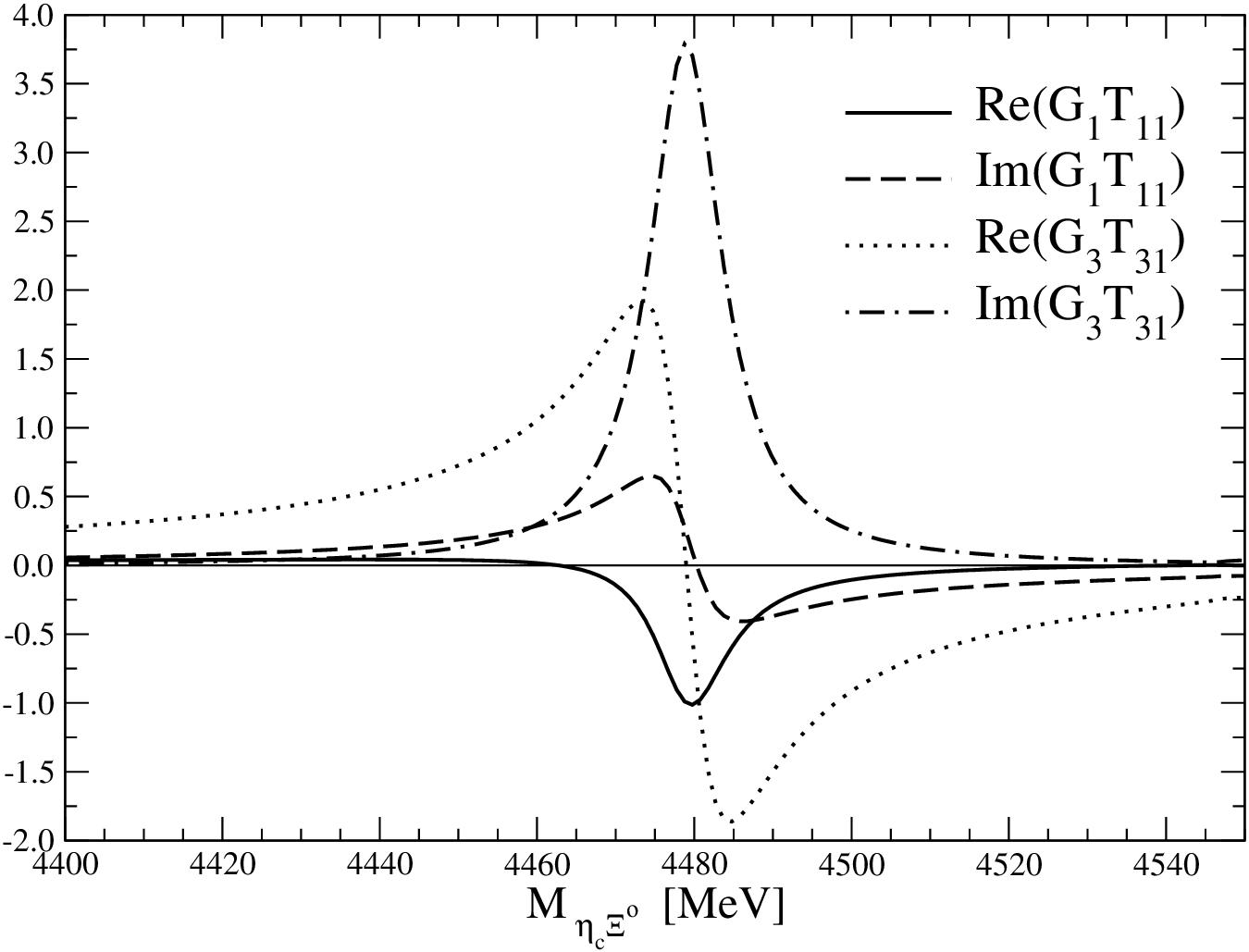}
    \caption{$GT$ terms contributing to the final state interaction using $\Lambda=800$~MeV. The index 1 stands for $\eta_c \Xi^0$ and 3 for $\bar{D}\Omega_c$. }
    \label{fig:GT}
\end{figure}
In addition, we see in Fig.~\ref{fig:GT} that the factor $G_3T_{31}\equiv G_{\bar D^0\Omega_c}\,t_{\bar D^0\Omega_c,\eta_c \Xi^0}$ in Eq.~\eqref{eqn:fullamplitudeMee} is typically about four times larger than the factor 
$G_1T_{11}\equiv G_{\eta_c \Xi^0}\,t_{\eta_c \Xi^0,\eta_c \Xi^0}$ in Eq.~\eqref{eqn:fullamplitude}, which contributes an extra factor of 16 in the distribution in favor also of the $\Omega_b^-$ decay. Combining this with the previously mentioned factor of 50 results in an 800-fold enhancement of the $\Omega_b^-$ spectrum compared to that of the $\Xi_b^0$.
On the other hand, the reason for the strong dip seen in the  $\Xi_b^0$ distribution is the interference between the FSI, Fig.~\ref{fig:diag_FSI}(b), and the tree level process, Fig.~\ref{fig:diag_FSI}(a). Mathematically, the tree level is accounted for by the 1 term in  Eq.~\eqref{eqn:fullamplitude},
which interferes with the $G_1T_{11}$ term from the FSI in that equation.
It can be seen in Fig.~\ref{fig:GT} that at the resonant position, the real part of the $G_1T_{11}$ term coincidentally takes a value very close to $-1$, leading to a destructive interference with the $+1$ term of the tree level mechanism.
This dip phenomenon is a well-known feature in scattering theory, typically arising from strong interference between a resonance and a non-resonant background.
In fact, if the tree-level contribution was absent, the spectrum for $\Xi_b^0$ decay would closely resemble that of $\Omega_b^-$ decay, except for the numerical factor $800$ explained above.

Summarizing the previous discussion, we can conclude that the $\eta_c \Xi^0$ spectrum in the $\Omega_b^-\to K^- \eta_c \Xi^0$ exhibits a strength approximately 800 times larger than that in  $\Xi_b^0\to \eta \eta_c \Xi^0$. Additionally, the $P_{css}$ state considered in the present work would appear as a Breit-Wigner-like shape in the former decay, while in the latter, it would manifest as a significant dip.

The decay modes under study are chosen as those that can provide the best access to the $P_{css}$ states. Another important factor is whether or not these decay processes can be reconstructed and analysed at current or future facilities. The requirement to study beauty baryon decays limits this discussion, at present, to the current and future upgrades of the LHCb experiment, where the proton-proton collisions provide access to all beauty hadrons. Using the Run~1 and Run~2 datasets, several decay modes of the $\Xi_b^{(0,-)}$ and $\Omega_b^-$ baryons have been observed~\cite{LHCb:2013jih,LHCb:2016hha,LHCb:2017gjj,LHCb:2023qxn,LHCb:2023ngz,LHCb:2024jqy}. However, the sample sizes from Run 1 and Run 2 are unlikely to be sufficient to observe the decays in question and certainly not large enough to enable amplitude analyses, where $\mathcal{O}(1000)$ candidates are required. These observations are of processes that are well-suited to the strengths of the LHCb experiment, reconstruction of charged tracks of hadrons and muons. We now consider the relative strengths and weaknesses for the decay processes considered here to be detected and reconstructed by the LHCb experiment. 

\begin{itemize}
\item{$\Xi_b^0\to \eta \eta_c \Xi^0$: This process will be challenging to reconstruct at LHCb due to the presence of neutral particles in the decays of the $\eta$ (e.g.\ $\gamma\gamma$ or $\pi^+\pi^-\pi^0$) and $\Xi^0 (\to \Lambda^0\pi^0)$ mesons. The relatively long-lived $\Lambda^0$ baryon introduces a secondary challenge because it will often decay downstream of the LHCb vertex locator. Such candidates are reconstructed using the downstream tracking detectors only with reduced efficiency. Finally, the $\eta_c$ meson appears in all three final states and can be reconstructed in hadronic final states such as $p\bar{p}$.}
\item{$\Omega_b^-\to K^- \eta_c \Xi^0$: Compared to the $\Xi_b^0$ decay mode above, the yield of $\Omega_b^-$ baryon decays will be suppressed by the ratio of production fractions in the $pp$ collisions at the LHC. However, swapping the $\eta$ meson for a charged kaon brings a large improvement in the reconstruction efficiency and will likely mean that this process is more promising to study first.}
\item{$\Omega_b^-\to K_S^0 \eta_c \Xi^-$:
While we have studied the decay $\Omega_b^-\to K^- \eta_c \Xi^0$, the $\Omega_b^-\to \bar K^0 \eta_c \Xi^-$ mode, to be seen experimentally in $\Omega_b^-\to K_S^0 \eta_c \Xi^-$, which has the same strength in our approach due to isospin symmetry, seems the most promising channel to study at the LHCb experiment. Reconstructing this final state requires no neutral particles because the $K_S^0 \to \pi^+\pi^-$ and $\Xi^-\to\Lambda^0\pi^-$ decay modes can be utilised. Similar to the $\Lambda^0$ baryon, the $K_S^0$ meson reconstruction efficiency is lower than for a charged kaon given the two-track final state and long lifetime that means many candidates will decay downstream of the LHCb vertex detector. Nevertheless, considering all of the particles involved, this process should be the most straightforward to study using LHCb data samples.}
\end{itemize}

It should also be noted that the branching fractions of all three decay modes are unknown, but given the quark-level processes involved, it is reasonable to expect they are all of similar magnitude and will not be significantly suppressed.

\section{Conclusions}

In light of previous theoretical predictions of several pentaquark states $P_{css}$, with global flavor  $\bar c c s s n$, based on formalisms implementing unitary in coupled channels, we have studied the feasibility of seeing the anticipated
$I(J^P)=\frac{1}{2}(\frac{1}{2}^-)$  pentaquark in the $\eta_c \Xi^0$ spectrum in the decays $\Xi_b^0\to \eta \eta_c \Xi^0$ and $\Omega_b^-\to K^- \eta_c \Xi^0$.
The theoretical model employed to generate the pentaquark  is based on techniques similar  to the chiral unitary approach, aiming to implement unitarity in coupled channels into the meson-baryon scattering amplitudes. The only input of the unitarization procedure is the tree level pseudoscalar-baryon potentials obtained from vector meson exchange mechanisms with Lagrangians derived from appropriate extensions of the local hidden gauge approach to the heavy quark sector.
The pentaquark states  emerge dynamically as poles in unphysical Riemann sheets of the unitarized pseudoscalar-baryon amplitudes, without the necessity of including them as explicit degrees of freedom.

In the present work, the pentaquarks are generated in the $\Xi_b^0$ and $\Omega_b^-$ decays by implementing the $\eta_c \Xi^0$ and
$\bar D^0 \Omega_c$
final state interaction, considering all the possible pseudoscalar-baryon coupled channels significant to generate the pentaquark according to the model. To evaluate the tree level $\Xi_b^0$ and $\Omega_b^-$ decay amplitudes, necessary before implementing the final state interaction, we identify the dominant mechanisms at the quark level. Subsequently, we implement the hadronization, after the weak transition, incorporating the generation of $\bar q q$ pairs within the philosophy of the $^3P_0$ model. The different channels are related by flavor symmetry considerations at the quark level.

We provide,  up to a global factor,  the $\eta_c \Xi^0$ invariant mass distributions for the 
$\Xi_b^0\to \eta \eta_c \Xi^0$ and $\Omega_b^-\to K^- \eta_c \Xi^0$ decays which manifest a large difference in both strength and shape between the two decays. The strength notably favors  the  $\Omega_b^-$ decay by about a factor 800, primarily influenced by flavor and color numerical coefficients in the elementary amplitude. Furthermore, the spectrum for the  $\Xi_b^0$ decay manifests the resonance as a distinctive dip in its shape, resulting from the interference between non-resonant tree level production and the resonant final state interaction responsible for the pentaquark generation. On the other hand, the  $\Omega_b^-$ decay leads to a very clear peak resembling a Breit-Wigner distribution.

Therefore, for both theoretical and experimental reasons, we recommend searches using the Run~3 and Run~4 datasets from LHCb to focus initially on searching for $\Omega_b^-\to K_S^0 \eta_c \Xi^-$ decays and studying the different contributing amplitudes to search for the $P_{css}$ states. Following this, the focus should first switch to study $\Omega_b^-\to K^- \eta_c \Xi^0$ decays and then finally to the $\Xi_b^0\to \eta \eta_c \Xi^0$ process, possibly requiring data samples from the proposed LHCb Upgrade 2 project~\cite{LHCb:2018roe}.

\section*{ACKNOWLEDGEMENTS}

This
work is partly supported by the Spanish Ministerio
de Economia y Competitividad (MINECO) and European FEDER funds under
Contracts No. FIS2017-84038-
C2-1-P B, PID2020-112777GB-I00, and by Generalitat
Valenciana under contract PROMETEO/2020/023. This
project has received funding from the European Union
Horizon 2020 research and innovation programme under
the program H2020-INFRAIA-2018-1, grant agreement
No. 824093 of the STRONG-2020 project.
This work is partly supported by the Science and Technology Facilities Council (UK).


\end{document}